# Software Engineering at Google

*Fergus Henderson*

*<fergus@google.com> (work) or*
*<fergus.henderson@gmail.com> (personal)*

## Abstract

We catalog and describe Google’s key software engineering practices.

## Biography

Fergus Henderson has been a software engineer at Google for over 10 years. He started programming as a kid in 1979, and went on to academic research in programming language design and implementation. With his PhD supervisor, he co-founded a research group at the University of Melbourne that developed the programming language Mercury. He has been a program committee member for eight international conferences, and has released over 500,000 lines of open-source code. He was a former moderator of the Usenet newsgroup comp.std.c++ and was an officially accredited “Technical Expert” to the ISO C and C++ committees. He has over 15 years of commercial software industry experience. At Google, he was one of the original developers of Blaze, a build tool now used across Google, and worked on the server-side software behind speech recognition and voice actions (before Siri!) and speech synthesis. He currently manages Google's text-to-speech engineering team, but still writes and reviews plenty of code. Software that he has written is installed on over a billion devices, and gets used over a billion times per day.

# 1. Introduction

**Google has been a phenomenally successful company**. As well as the success of Google Search and AdWords, Google has delivered many other stand-out products, including Google Maps, Google News, Google Translate, Google speech recognition, Chrome, and Android. Google has also greatly enhanced and scaled many products that were acquired by purchasing smaller companies, such as YouTube, and has made significant contributions to a wide variety of open-source projects. And Google has demonstrated some amazing products that are yet to launch, such as self-driving cars.

There are many reasons for Google's success, including enlightened leadership, great people, a high hiring bar, and the financial strength that comes from successfully taking advantage of an early lead in a very rapidly growing market. But one of these reasons is that **Google has developed excellent software engineering practices**, which have helped it to succeed. These practices have evolved over time based on the accumulated and distilled wisdom of many of the most talented software engineers on the planet. We would like to share knowledge of our practices with the world, and to share some of the lessons that we have learned from our mistakes along the way.

**The aim of this paper is to catalogue and briefly describe Google's key software engineering practices.** Other organizations and individuals can then compare and contrast these with their own software engineering practices, and consider whether to apply some of these practices themselves.

Many authors (e.g. [9], [10], [11]) have written books or articles analyzing Google's success and history. But most of those have dealt mainly with business, management, and culture; only a fraction of those (e.g. [1, 2, 3, 4, 5, 6, 7, 13, 14, 16, 21]) have explored the software engineering side of things, and most explore only a single aspect; and none of them provide a brief written overview of software engineering practices at Google as a whole, as this paper aims to do.

# 2. Software development

## 2.1. The Source Repository

**Most of Google's code is stored in a single unified source-code repository, and is accessible to all software engineers at Google**. (Customer data is tightly secured, it's only source code that's accessible.) There are some notable exceptions to the use of this single widely accessible repository, particularly the two large open-source projects Chrome and Android, which use separate open-source repositories, and some high-value or security-critical pieces of code for which read access is locked down more tightly. But most Google projects

share the same repository. As of January 2015, this 86 terabyte repository contained a billion files, including over 9 million source code files containing a total of **2 billion lines of source code**, with a history of 35 million commits and a change rate of 40 thousand commits per work day [18]. Write access to the repository is controlled: only the listed owners of each subtree of the repository can approve changes to that subtree. But generally any engineer can access any piece of code, can check it out and build it, can make local modifications, can test them, and can send changes for review by the code owners, and if an owner approves, can check in (commit) those changes. Culturally, engineers are encouraged to fix anything that they see is broken and know how to fix, regardless of project boundaries. This empowers engineers and leads to higher-quality infrastructure that better meets the needs of those using it.

**Almost all development occurs at the "head" of the repository**, not on branches. This helps identify integration problems early and minimizes the amount of merging work needed. It also makes it much easier and faster to push out security fixes.

**Automated systems run tests frequently,** often after every change to any file in the transitive dependencies of the test, although this is not always feasible. These systems automatically notify the author and reviewers of any change for which the tests failed, typically within a few minutes. Most teams make the current status of their build very conspicuous by installing prominent displays or even sculptures with color-coded lights (green for building successfully and all tests passing, red for some tests failing, black for broken build). This helps to focus engineers' attention on keeping the build green. Most larger teams also have a "build cop" who is responsible for ensuring that the tests continue to pass at head, by working with the authors of the offending changes to quickly fix any problems or to roll back the offending change. (The build cop role is typically rotated among the team or among its more experienced members.) This focus on keeping the build green makes development at head practical, even for very large teams.

**Code ownership.** Each subtree of the repository can have a file listing the user ids of the "owners" of that subtree. Subdirectories also inherit owners from their parent directories, although that can be optionally suppressed. The owners of each subtree control write access to that subtree, as described in the code review section below. Each subtree is required to have at least two owners, although typically there are more, especially in geographically distributed teams. It is common for the whole team to be listed in the owners file. Changes to a subtree can be made by anyone at Google, not just the owners, but must be approved by an owner. This ensures that every change is reviewed by an engineer who understands the software being modified.

For more on the source code repository at Google, see [17, 18, 21]; and for how another large company deals with the same challenge, see [19].

## 2.2. The Build System

Google uses a distributed build system known as Blaze, which is responsible for compiling and linking software and for running tests. It provides standard commands for building and testing software that work across the whole repository. These standard commands and the highly optimized implementation mean that **it is typically very simple and quick for any Google engineer to build and test any software in the repository**. This consistency is a key enabler which helps to make it practical for engineers to make changes across project boundaries.

Programmers write "BUILD" files that Blaze uses to determine how to build their software. Build entities such as libraries, programs, and tests are declared using fairly high-level **declarative build specifications** that specify, for each entity, its name, its source files, and the libraries or other build entities that it depends on. These build specifications are comprised of declarations called "build rules" that each specify high-level concepts like "here is a C++ library with these source files which depends on these other libraries", and it is up to the build system to map each build rule to a set of build steps, e.g. steps for compiling each source file and steps for linking, and for determining which compiler and compilation flags to use.

In some cases, notably Go programs, build files can be generated (and updated) automatically, since the dependency information in the BUILD files is (often) an abstraction of the dependency information in the source files. But they are nevertheless checked in to the repository. This ensures that the build system can quickly determine dependencies by analyzing only the build files rather than the source files, and it avoids excessive coupling between the build system and compilers or analysis tools for the many different programming languages supported.

The build system's implementation uses Google's distributed computing infrastructure. The work of each build is typically **distributed across hundreds or even thousands of machines**. This makes it possible to build extremely large programs quickly or to run thousands of tests in parallel.

**Individual build steps must be "hermetic": they depend only on their declared inputs.** Enforcing that all dependencies be correctly declared is a consequence of distributing the build: only the declared inputs are sent to the machine on which the build step is run. As a result the build system can be relied on to know the true dependencies. Even the compilers that the build system invokes are treated as inputs.

**Individual build steps are deterministic.** As a consequence, the build system can cache build results. Software engineers can sync their workspace back to an old change number and can rebuild and will get exactly the same binary. Furthermore, this cache can be safely shared between different users. (To make this work properly, we had to eliminate non-determinism in the tools invoked by the build, for example by scrubbing out timestamps in the generated output files.)

**The build system is reliable.** The build system tracks dependencies on changes to the build rules themselves, and knows to rebuild targets if the action to produce them changed, even if the inputs to that action didn't, for example when only the compiler options changed. It also deals properly with interrupting the build part way, or modifying source files during the build: in such cases, you need only rerun the build command. There is never any need to run the equivalent of "make clean".

**Build results are cached "in the cloud"**. This includes intermediate results. If another build request needs the same results, the build system will automatically reuse them rather than rebuilding, even if the request comes from a different user.

**Incremental rebuilds are fast.** The build system stays resident in memory so that for rebuilds it can incrementally analyze just the files that have changed since the last build.

**Presubmit checks.** Google has tools for automatically running a suite of tests when initiating a code review and/or preparing to commit a change to the repository. Each subtree of the repository can contain a configuration file which determines which tests to run, and whether to run them at code review time, or immediately before submitting, or both. The tests can be either synchronous, i.e. run before sending the change for review and/or before committing the change to the repository (good for fast-running tests); or asynchronous, with the results emailed to the review discussion thread. [The review thread is the email thread on which the code review takes place; all the information in that thread is also displayed in the web-based code review tool.]

## 2.3. Code Review

**Google has built excellent web-based code review tools**, integrated with email, that allow authors to request a review, and allows reviewers to view side-by-side diffs (with nice color coding) and comment on them. When the author of a change initiates a code review, the reviewers are notified by e-mail, with a link to the web review tool's page for that change. Email notifications are sent when reviewers submit their review comments. In addition, automated tools can send notifications, containing for example the results of automated tests or the findings of static analysis tools.

**All changes to the main source code repository MUST be reviewed by at least one other engineer.** In addition, if the author of a change is not one of the owners of the files being modified, then at least one of the owners must review and approve the change.

In exceptional cases, an owner of a subtree can check in (commit) an urgent change to that subtree *before* it is reviewed, but a reviewer must still be named, and the change author and reviewer will get automatically nagged about it until the change has been reviewed and approved. In such cases, any modifications needed to address review comments must be done in a separate change, since the original change will have already been committed.

Google has tools for automatically suggesting reviewer(s) for a given change, by looking at the ownership and authorship of the code being modified, the history of recent reviewers, and the number of pending code reviews for each potential reviewer. At least one of the owners of each subtree which a change affects must review and approve that change. But apart from that, the author is free to choose reviewer(s) as they see fit.

One potential issue with code review is that if the reviewers are too slow to respond or are overly reluctant to approve changes, this could potentially slow down development. The fact that the code author chooses their reviewers helps avoid such problems, allowing engineers to avoid reviewers that might be overly possessive about their code, or to send reviews for simple changes to less thorough reviewers and to send reviews for more complex changes to more experienced reviewers or to several reviewers.

**Code review discussions for each project are automatically copied to a mailing list designated by the project maintainers.** Anyone is free to comment on any change, regardless of whether they were named as a reviewer of that change, both before and after the change is committed. If a bug is discovered, it’s common to track down the change that introduced it and to comment on the original code review thread to point out the mistake so that the original author and reviewers are aware of it.

It is also possible to send code reviews to several reviewers and then to commit the change as soon as one of them has approved (provided either the author or the first responding reviewer is an owner, of course), before the other reviewers have commented, with any subsequent review comments being dealt with in follow-up changes. This can reduce the turnaround time for reviews.

In addition to the main section of the repository, **there is an “experimental” section of the repository where the normal code review requirements are not enforced**. However, code running in production must be in the main section of the repository, and engineers are very strongly encouraged to develop code in the main section of the repository, rather than developing in experimental and then moving it to the main section, since code review is most effective when done as the code is developed rather than afterwards. In practice engineers often request code reviews even for code in experimental.

**Engineers are encouraged to keep each individual change small**, with larger changes preferably broken into a series of smaller changes that a reviewer can easily review in one go. This also makes it easier for the author to respond to major changes suggested during the review of each piece; very large changes are often too rigid and resist reviewer-suggested changes. One way in which keeping changes small is encouraged[1] is that the code review tools label each code review with a description of the size of the change, with changes of 30-99 lines added/deleted/removed being labelled “medium-size” and with changes of above 300 lines being labelled with increasingly disparaging labels, e.g. “large” (300-999), “freakin huge” (1000-1999), etc. (However, in a typically Googly way, this is kept fun by replacing these familiar descriptions with amusing alternatives on a few days each year, such as talk-like-a-pirate day. :)

## 2.4. Testing

**Unit Testing is strongly encouraged and widely practiced at Google**. All code used in production is expected to have unit tests, and the code review tool will highlight if source files are added without corresponding tests. Code reviewers usually require that any change which adds new functionality should also add new tests to cover the new functionality. Mocking frameworks (which allow construction of lightweight unit tests even for code with dependencies on heavyweight libraries) are quite popular.

Integration testing and regression testing are also widely practiced.

As discussed in "Presubmit Checks" above, testing can be automatically enforced as part of the code review and commit process.

Google also has automated tools for measuring test coverage. The results are also integrated as an optional layer in the source code browser.

**Load testing prior to deployment** is also de rigueur at Google. Teams are expected to produce a table or graph showing how key metrics, particularly latency and error rate, vary with the rate of incoming requests.

## 2.5. Bug tracking

Google uses a bug tracking system called Buganizer for tracking issues: bugs, feature requests, customer issues, and processes (such as releases or clean-up efforts). Bugs are categorized into hierarchical components and each component can have a default assignee and default email list to CC. When sending a source change for review, engineers are prompted to associate the change with a particular issue number.

It is common (though not universal) for teams at Google to regularly scan through open issues in their component(s), prioritizing them and where appropriate assigning them to particular engineers. Some teams have a particular individual responsible for bug triage, others do bug triage in their regular team meetings. Many teams at Google make use of labels on bugs to indicate whether bugs have been triaged, and which release(s) each bug is targeted to be fixed in.

---

[1] This has changed somewhat in recent years. More recent versions of the code review tools no longer use the more disparaging labels for large CLs, but they are still labelled with their size, e.g. “S”, “M”, “L”, “XL”.

## 2.6. Programming languages

Software engineers at Google are strongly encouraged to program in one of five officially-approved programming languages at Google: **C++, Java, Python, Go, or JavaScript**. Minimizing the number of different programming languages used reduces obstacles to code reuse and programmer collaboration.

There are also Google **style guides** for each language, to ensure that code all across the company is written with similar style, layout, naming conventions, etc. In addition there is a company-wide **readability** training process, whereby experienced engineers who care about code readability train other engineers in how to write readable, idiomatic code in a particular language, by reviewing a substantial change or series of changes until the reviewer is satisfied that the author knows how to write readable code in that language. Each change that adds non-trivial new code in a particular language must be approved by someone who has passed this “readability” training process in that language.

In addition to these five languages, many **specialized domain-specific languages** are used for particular purposes (e.g. the build language used for specifying build targets and their dependencies).

Interoperation between these different programming languages is done mainly using **Protocol Buffers**. Protocol Buffers is a way of encoding structured data in an efficient yet extensible way. It includes a domain-specific language for specifying structured data, together with a compiler that takes in such descriptions and generates code in C++, Java, Python, for constructing, accessing, serializing, and deserializing these objects. Google’s version of Protocol Buffers is integrated with Google’s RPC libraries, enabling simple cross-language RPCs, with serialization and deserialization of requests and responses handled automatically by the RPC framework.

**Commonality of process** is a key to making development easy even with an enormous code base and a diversity of languages: there is a single set of commands to perform all the usual software engineering tasks (such as check out, edit, build, test, review, commit, file bug report, etc.) and the same commands can be used no matter what project or language. Developers don’t need to learn a new development process just because the code that they are editing happens to be part of a different project or written in a different language.

## 2.7. Debugging and Profiling tools

Google servers are linked with libraries that provide a number of tools for debugging running servers. In case of a server crash, a signal handler will automatically dump a stack trace to a log file, as well as saving the core file. If the crash was due to running out of heap memory, the server will dump stack traces of the allocation sites of a sampled subset of the live heap objects. There are also web interfaces for debugging that allow examining incoming and outgoing RPCs (including timing, error rates, rate limiting, etc.), changing command-line flag values (e.g. to increase logging verbosity for a particular module), resource consumption, profiling, and more. These tools greatly increase the overall ease of debugging to the point where it is rare to fire up a traditional debugger such as gdb.

## 2.8. Release engineering

A few teams have dedicated release engineers, but for most teams at Google, the release engineering work is done by regular software engineers.

**Releases are done frequently** for most software; weekly or fortnightly releases are a common goal, and some teams even release daily. This is made possible by **automating most of the normal release engineering tasks**. Releasing frequently helps to keep engineers motivated (it's harder to get excited about something if it won't be released until many months or even years into the future) and increases overall velocity by allowing more iterations, and thus more opportunities for feedback and more chances to respond to feedback, in a given time.

A release typically starts in a fresh workspace, by syncing to the change number of the latest "green" build (i.e. the last change for which all the automatic tests passed), and making a release branch. The release engineer can select additional changes to be "cherry-picked", i.e. merged from the main branch onto the release branch. Then the software will be rebuilt from scratch and the tests are run. If any tests fail, additional changes are made to fix the failures and those additional changes are cherry-picked onto the release branch, after which the software will be rebuilt and the tests rerun. When the tests all pass, the built executable(s) and data file(s) are packaged up. All of these steps are automated so that the release engineer need only run some simple commands, or even just select some entries on a menu-driven UI, and choose which changes (if any) to cherry pick.

Once a candidate build has been packaged up, it is typically loaded onto a "**staging**" server for further **integration testing by small set of users** (sometimes just the development team).

A useful technique involves sending a copy of (a subset of) the requests from production traffic to the staging server, but also sending those same requests to the current production servers for actual processing. The responses from the staging server are discarded, and the responses from the live production servers are sent back to the users. This helps ensure that any issues that might cause serious problems (e.g. server crashes) can be detected before putting the server into production.

The next step is to usually roll out to one or more "**canary**" servers that are **processing a subset of the live production traffic.** Unlike the "staging" servers, these are processing and responding to real users.

Finally the release can be rolled out to all servers in all data centers. For very high-traffic, high-reliability services, this is done with a **gradual roll-out** over a period of a couple of days, to help reduce the impact of any outages due to newly introduced bugs not caught by any of the previous steps.

For more information on release engineering at Google, see chapter 8 of the SRE book [7]. See also [15].

## 2.9. Launch approval

The launch of any user-visible change or significant design change requires approvals from a number of people outside of the core engineering team that implements the change. In particular approvals (often subject to detailed review) are required to ensure that code complies with legal requirements, privacy requirements, security requirements, reliability requirements (e.g. having appropriate automatic monitoring to detect server outages and automatically notify the appropriate engineers), business requirements, and so forth.

The launch process is also designed to ensure that appropriate people within the company are notified whenever any significant new product or feature launches.

**Google has an internal launch approval tool that is used to track the required reviews and approvals and ensure compliance with the defined launch processes for each product.** This tool is easily customizable, so that different products or product areas can have different sets of required reviews and approvals.

For more information about launch processes, see chapter 27 of the SRE book [7].

## 2.10. Post-mortems

Whenever there is a significant outage of any of our production systems, or similar mishap, the people involved are required to write a post-mortem document. This document describes the incident, including title, summary, impact, timeline, root cause(s), what worked/what didn't, and action items. **The focus is on the problems, and how to avoid them in future, not on the people or apportioning blame.** The impact section tries to quantify the effect of the incident, in terms of duration of outage, number of lost queries (or failed RPCs, etc.), and revenue. The timeline section gives a timeline of the events leading up to the outage and the steps taken to diagnose and rectify it. The what worked/what didn't section describes the lessons learnt -- which practices helped to quickly detect and resolve the issue, what went wrong, and what concrete actions (preferably filed as bugs assigned to specific people) can be take to reduce the likelihood and/or severity of similar problems in future.

For more information on post-mortem culture at Google, see chapter 15 of the SRE book [7].

### 2.11. Frequent rewrites

**Most software at Google gets rewritten every few years.**

This may seem incredibly costly. Indeed, it does consume a large fraction of Google's resources. However, it also has some crucial benefits that are key to Google's agility and long-term success. In a period of a few years, it is typical for the requirements for a product to change significantly, as the software environment and other technology around it change, and as changes in technology or in the marketplace affect user needs, desires, and expectations. Software that is a few years old was designed around an older set of requirements and is typically not designed in a way that is optimal for current requirements. Furthermore, it has typically accumulated a lot of complexity. Rewriting code cuts away all the unnecessary accumulated complexity that was addressing requirements which are no longer so important. In addition, rewriting code is a way of transferring knowledge and a sense of ownership to newer team members. This sense of ownership is crucial for productivity: engineers naturally put more effort into developing features and fixing problems in code that they feel is "theirs". Frequent rewrites also encourage mobility of engineers between different projects which helps to encourage cross-pollination of ideas. Frequent rewrites also help to ensure that code is written using modern technology and methodology.

## 3. Project management

### 3.1. 20% time

**Engineers are permitted to spend up to 20% of their time working on any project of their choice, without needing approval from their manager or anyone else.** This trust in engineers is extremely valuable, for several reasons. Firstly, it allows anyone with a good idea, even if it is an idea that others would not immediately recognize as being worthwhile, to have sufficient time to develop a prototype, demo, or presentation to show the value of their idea. Secondly, it provides management with visibility into activity that might otherwise be hidden. In other companies that don't have an official policy of allowing 20% time, engineers sometimes work on "skunkwork" projects without informing management. It's much better if engineers can be open about such projects, describing their work on such projects in their regular status updates, even in cases where their management may not agree on the value of the project. Having a company-wide official policy and a culture that supports it makes this possible. Thirdly, by allowing engineers to spend a small portion of their time working on more fun stuff, it keeps engineers motivated and excited by what they do, and stops them getting burnt out, which can easily happen if they feel compelled to spend 100% of their time working on more tedious tasks. The difference in productivity between engaged, motivated engineers and burnt out engineers is

a *lot* more than 20%. Fourthly, it encourages a culture of innovation. Seeing other engineers working on fun experimental 20% projects encourages everyone to do the same.

## 3.2. Objectives and Key Results (OKRs)

**Individuals and teams at Google are required to explicitly document their goals and to assess their progress towards these goals.** Teams set quarterly and annual objectives, with measurable key results that show progress towards these objectives. This is done at every level of the company, going all the way up to defining goals for the whole company. Goals for individuals and small teams should align with the higher-level goals for the broader teams that they are part of and with the overall company goals. At the end of each quarter, progress towards the measurable key results is recorded and each objective is given a score from 0.0 (no progress) to 1.0 (100% completion). OKRs and OKR scores are normally made visible across Google (with occasional exceptions for especially sensitive information such as highly confidential projects), but they *not* used directly as input to an individual's performance appraisal.

OKRs should be set high: the desired target overall average score is 65%, meaning that a team is encouraged to set as goals about 50% more tasks than they are likely to actually accomplish. If a team scores significantly higher than that, they are encouraged to set more ambitious OKRs for the following quarter (and conversely if they score significantly lower than that, they are encouraged to set their OKRs more conservatively the next quarter).

OKRs provide a key mechanism for communicating what each part of the company is working on, and for encouraging good performance from employees via social incentives… engineers know that their team will have a meeting where the OKRs will be scored, and have a natural drive to try to score well, even though OKRs have no direct impact on performance appraisals or compensation. Defining key results that are objective and measurable helps ensure that this human drive to perform well is channelled to doing things that have real concrete measurable impact on progress towards shared objectives.

## 3.3. Project approval

Although there is a well-defined process for launch approvals, Google does not have a well-defined process for project approval or cancellation. Despite having been at Google for over 10 years, and now having become a manager myself, I still don't fully understand how such decisions are made. In part this is because the approach to this is not uniform across the company. Managers at every level are responsible and accountable for what projects their teams work on, and exercise their discretion as they see fit. In some cases, this means that such decisions are made in a quite bottom-up fashion, with engineers being given freedom to choose which projects to work on, within their team's scope. In other cases, such decisions are made in a much more top-down fashion, with executives or managers making decisions about which projects will go ahead, which will get additional resources, and which will get cancelled.

## 3.4. Corporate reorganizations

Occasionally an executive decision is made to cancel a large project, and then the many engineers who had been working on that project may have to find new projects on new teams. Similarly there have been occasional “defragmentation” efforts, where projects that are split across multiple geographic locations are consolidated into a smaller number of locations, with engineers in some locations being required to change team and/or project in order to achieve this. In such cases, engineers are generally given freedom to choose their new team and role from within the positions available in their geographic location, or in the case of defragmentation, they may also be given the option of staying on the same team and project by moving to a different location.

In addition, other kinds of corporate reorganizations, such as merging or splitting teams and changes in reporting chains, seem to be fairly frequent occurrences, although I don’t know how Google compares with other large companies on that. In a large, technology-driven organization, somewhat frequent reorganization may be necessary to avoid organizational inefficiencies as the technology and requirements change. They can help to avoid or mitigate the common problem in large organizations of “shipping the org chart”, where the software architecture reflects the organization’s reporting structure.

## 3.5. Annual hack weeks

Another way in which Google encourages innovation is by holding “hackathons” [23]. In many Google offices, this takes the form of an annual week-long “hackathon” in which software engineers can take time out of their regular schedule to work on new innovative projects. After a kick-off meeting in which people can pitch ideas, they form small teams that spend a week building a demo or prototype based on a new idea, and at the end of the week present their demos to an audience and a panel of judges, who award (small) prizes for the best projects. The biggest rewards, though, are the recognition — and the chance to have a successful hackathon project graduate to become a full-time real project.

Despite all engineers being given permission from their site lead to participate in such hackathons, many engineers do find it difficult to detach from their day-to-day responsibilities, and so participation rates are generally fairly low (e.g. 5-20%), although many more will attend the initial kick-off and/or the final presentations and may get inspired by the ideas presented.

# 4. People management

## 4.1. Roles

As we’ll explain in more detail below, Google separates the engineering and management career progression ladders, separates the tech lead role from management, embeds research within engineering, and supports engineers with product managers, project managers, and site reliability engineers (SREs). It seems likely that at least some of these practices are important to sustaining the culture of innovation that has developed at Google.

Google has a small number of different roles within engineering. Within each role, there is a career progression possible, with a sequence of levels, and the possibility of promotion (with associated improvement to compensation, e.g. salary) to recognize performance at the next level.

The main roles are these:

- **Engineering Manager**

  This is the only people management role in this list. Individuals in other roles such as Software Engineer *may* manage people, but Engineering Managers *always* manage people. Engineering Managers are often former Software Engineers, and invariably have considerable technical expertise, as well as people skills.

  **There is a distinction between technical leadership and people management.** Engineering Managers do not necessarily lead projects; projects are led by a Tech Lead, who can be an Engineering Manager, but who is more often a Software Engineer. A project's Tech Lead has the final say for technical decisions in that project.

  Managers are responsible for selecting Tech Leads, and for the performance of their teams. They perform coaching and assisting with career development, do performance evaluation (using input from peer feedback, see below), and are responsible for some aspects of compensation. They are also responsible for some parts of the hiring process.

  Engineering Managers normally directly manage anywhere between 3 and 30 people, although 8 to 12 is most common.

- **Software Engineer (SWE)**

  Most people doing software development work have this role. The hiring bar for software engineers at Google is very high; by hiring only exceptionally good software engineers, a lot of the software problems that plague other organizations are avoided or minimized.

  Like many modern software companies, **Google has separate career progression sequences for engineering and management**. Although it is possible for a Software Engineer to manage people, or to transfer to the Engineering Manager role, managing people is *not* a requirement for promotion, even at the highest levels. At the higher levels, showing leadership is required, but that can come in many forms. For example creating great software that has a huge impact or is used by very many other engineers is sufficient. This is important, because it means that people who have great technical skills but lack the desire or skills to manage people still have a good career progression path that does not require them to take a management track. This avoids the problem that some organizations suffer where people end up in management positions for reasons of career advancement but neglect the people management of the people in their team.

- **Research Scientist**

  The hiring criteria for this role are very strict, and the bar is extremely high, requiring demonstrated exceptional research ability evidenced by a great publication record *and* ability to write code. Many very talented people in academia who would be able to qualify for a Software Engineer role would not qualify for a Research Scientist role at Google; most of the people with PhDs at Google are Software Engineers rather than Research Scientists. Research scientists are evaluated on their research contributions, including their publications, but apart from that and the different title, there is not really that much difference between the Software Engineer and Research Scientist role at Google. Both can do original research and publish papers, both can develop new product ideas and new technologies, and both can and do write code and develop products. Research Scientists at Google usually work alongside Software Engineers, in the same teams and working on the same products or the same research. This practice of embedding research within engineering contributes greatly to the ease with which new research can be incorporated into shipping products.

- **Site Reliability Engineer (SRE)**

  The maintenance of operational systems is done by software engineering teams, rather than traditional sysadmin types, but the hiring requirements for SREs are slightly different than the requirements for the Software Engineer position (software engineering skills requirements can be slightly lower, if compensated for by expertise in other skills such as networking or unix system internals). The nature and purpose of the SRE role is explained very well and in detail in the SRE book [7], so we won't discuss it further here.

- **Product Manager**

  Product Managers are responsible for the management of a product; as advocates for the product users, they coordinate the work of software engineers, evangelizing features of importance to those users, coordinating with other teams, tracking bugs and schedules, and ensuring that everything needed is in place to produce a high quality product. Product Managers usually do NOT write code themselves, but work with software engineers to ensure that the right code gets written.

- **Program Manager / Technical Program Manager**

  Program Managers have a role that is broadly similar to Product Manager, but rather than managing a product, they manage projects, processes, or operations (e.g. data collection). Technical Program Managers are similar, but also require specific technical expertise relating to their work, e.g. linguistics for dealing with speech data.

  The ratio of Software Engineers to Product Managers and Program Managers varies across the organization, but is generally high, e.g. in the range 4:1 to 30:1.

## 4.2. Facilities

Google is famous for its fun facilities, with features like slides, ball pits, and games rooms. That helps attract and retain good talent. Google's excellent cafes, which are free to employees, provide that function too, and also subtly encourage Googlers to stay in the office; hunger is never a reason to leave. The frequent placement of "microkitchens" where employees can grab snacks and drinks serves the same function too, but also acts as an important source of informal idea exchange, as many conversations start up there. Gyms, sports, and on-site massage help keep employees fit, healthy, and happy, which improves productivity and retention.

The seating at Google is open-plan, and often fairly dense. While controversial [20], this encourages communication, sometimes at the expense of individual concentration, and is economical.

Employees are assigned an individual seat, but seats are re-assigned fairly frequently (e.g. every 6-12 months, often as a consequence of the organization expanding), with seating chosen by managers to facilitate and encourage communication, which is always easier between adjacent or nearly adjacent individuals.

Google's facilities all have meeting rooms fitted with state-of-the-art video conference facilities, where connecting to the other party for a prescheduled calendar invite is just a single tap on the screen.

## 4.3. Training

Google encourages employee education in many ways:

- New Googlers ("Nooglers") have a mandatory initial training course.
- Technical staff (SWEs and research scientists) start by doing "Codelabs": short online training courses in individual technologies, with coding exercises.

- Google offers employees a variety of online and in-person training courses.
- Google also offers support for studying at external institutions.

In addition, each Noogler is usually appointed an official "Mentor" and a separate "Buddy" to help get them up to speed. Unofficial mentoring also occurs via regular meetings with their manager, team meetings, code reviews, design reviews, and informal processes.

## 4.4. Transfers

**Transfers between different parts of the company are encouraged**, to help spread knowledge and technology across the organization and improve cross-organization communication. Transfers between projects and/or offices are allowed for employees in good standing after 12 months in a position. Software engineers are also encouraged to do temporary assignments in other parts of the organization, e.g. a six-month "rotation" (temporary assignment) in SRE (Site Reliability Engineering).

## 4.5. Performance appraisal and rewards

Feedback is strongly encouraged at Google. Engineers can give each other explicit positive feedback via "peer bonuses" and "kudos". Any employee can nominate any other employee for a "peer bonus" -- a cash bonus of $100 -- up to twice per year, for going beyond the normal call of duty, just by filling in a web form to describe the reason. Team-mates are also typically notified when a peer bonus is awarded. Employees can also give "kudos", formalized statements of praise which provide explicit social recognition for good work, but with no financial reward; for "kudos" there is no requirement that the work be beyond the normal call of duty, and no limit on the number of times that they can be bestowed.

Managers can also award bonuses, including spot bonuses, e.g. for project completion. And as with many companies, Google employees get annual performance bonuses and equity awards based on their performance.

Google has a very careful and detailed promotion process, which involves nomination by self or manager, self-review, peer reviews, manager appraisals; the actual decisions are then made by promotion committees based on that input, and the results can be subject to further review by promotion appeals committees. Ensuring that the right people get promoted is critical to maintaining the right incentives for employees.

Poor performance, on the other hand, is handled with manager feedback, and if necessary with performance improvement plans, which involve setting very explicit concrete performance targets and assessing progress towards those targets. If that fails, termination for poor performance is possible, but in practice this is *extremely* rare at Google.

Manager performance is assessed with feedback surveys; every employee is asked to fill in an survey about the performance of their manager twice a year, and the results are anonymized and aggregated and then made available to managers. This kind of upward feedback is very important for maintaining and improving the quality of management throughout the organization.

# 5. Conclusions

We have briefly described most of the key software engineering practices used at Google. Of course Google is now a large and diverse organization, and some parts of the organization have different practices. But the practices described here are generally followed by most teams at Google.

With so many different software engineering practices involved, and with so many other reasons for Google's success that are not related to our software engineering practices, it is extremely difficult to give any quantitative or objective evidence connecting individual practices with improved outcomes. However, these practices are the ones that have stood the test of time at Google, where they have been subject to the collective subjective judgement of many thousands of excellent software engineers.

# Acknowledgements

Special thanks to Alan Donovan for his extremely detailed and constructive feedback, and thanks also to Yaroslav Volovich, Urs Hölzle, Brian Strope, Alexander Gutkin, Alex Gruenstein, Hameed Husaini, Glen Shires, Niall Murphy, Ranjit Mathew, Emma Soederberg, and Rob Siemborski for their very helpful comments on earlier drafts of this paper.